\documentclass[twocolumn,showpacs,preprintnumbers,amsmath,amssymb,prl,superscriptaddress]{revtex4}

\usepackage{graphicx}
\usepackage{dcolumn}
\usepackage{bm}

\begin{document}


\title{Superconductivity in a Molecular Metal Cluster Compound}

\author{O.N.~Bakharev}
\altaffiliation[Presently at: ]{Dept. of Chemistry, Univ. of
Aarhus, Denmark}
\author{D.~Bono}
\author{H.B.~Brom}
\affiliation{%
Kamerlingh Onnes Laboratory, Leiden University, P.O. Box 9504,
2300RA Leiden, The Netherlands
}%

\author{A.~Schnepf}
\author{H.~Schn\"ockel}
\affiliation{ Institut f\"ur Anorganische Chemie, Universit\"at
Karlsruhe, 76128 Karlsruhe, Germany
}%

\author{L.J.~de Jongh}%
\affiliation{%
Kamerlingh Onnes Laboratory, Leiden University, P.O. Box 9504,
2300RA Leiden, The Netherlands
}%

\date{\today}

\begin{abstract}
Compelling evidence for band-type conductivity and even bulk
superconductivity below $T_{\text{c}}\approx 8$~K has been found
in $^{69,71}$Ga-NMR experiments in crystalline ordered, giant
Ga$_{84}$ cluster-compounds. This material appears to represent
the first realization of a theoretical model proposed by Friedel
in 1992 for superconductivity in ordered arrays of weakly coupled,
identical metal nanoparticles.
\end{abstract}

\pacs{74.78.Na, 74.70.-b, 76.60.-k}
\keywords{Suggested keywords}

\maketitle

In recent years it has become apparent that the chemical
(bottom-up) route to nanostructures can be quite successful.
Molecular metal cluster compounds form an excellent example in
this respect \cite{Schmid98B}. These stoichiometric compounds form
macromolecular solids, in which the cores of the macromolecules
can be seen as metal nanoparticles. Often the cluster molecules
are ionic and, together with suitable counter-ions, form
crystalline 3D lattices. These ``self-organized nanostructures''
can thus be viewed as 3D-ordered arrays of identical metal
nanoparticles, embedded in the dielectric matrix formed by ligand
molecules plus counterions. Until recently, in all compounds
studied so far electron transfer between clusters proved
negligible, the materials being electrically insulating
\cite{Reedijk98,deJongh94}. Accordingly, the experiments were
probing single-particle properties at the nano-scale, such as
surface effects, quantum-size effects and the size-induced
metal-nonmetal transition \cite{Mulder94}.

However, given the strong similarity of metal cluster compounds to
e.g. the (alkali-doped) fullerenes (C$_{60}$), it was expected
\cite{deJongh94} that by doping or introducing mixed-valency, a
novel class of materials showing (super)conductivity could be
obtained. Indeed, a few years ago, the mixed-valent ``giant''
Ga$_{84}$ cluster compound
Ga$_{84}$[N(SiMe$_{3}$)$_{2}$]$_{20}$\--Li$_{6}$Br$_{2}$(thf)$_{20}\cdot$\-2toluene
was synthesized \cite{Schnepf01}, its crystal and molecular
structure being determined by X-ray diffraction \cite{Schnepf03}.
The [Ga$_{84}$R$_{20}$]$^{4-}$ molecules (called Ga$_{84}^{4-}$ in
what follows) form a fully ordered ionic crystal together with the
counterions (2[Li(thf)$_{4}$]$^{+}$ and
2[Li$_{2}$Br(thf)$_{6}$]$^{+}$). The mixed valent property arises
since the ion [Ga$_{84}$R$_{20}$]$^{3-}$ (Ga$_{84}^{3-}$) also
exists \cite{Schnepf03}, having the same molecular structure but
different crystalline packing of the cluster molecules
\cite{Schnepf03}. Since synthetic conditions for both moieties are
different, samples obtained are ``crystallographically pure'',
meaning that by X-ray no trace of the other moiety can be
detected. However, as evidenced by EPR, a small amount (~1\%) of
mixed valency is still present,probably due to local departures
from stoichiometry in the concentration of counterions.

In earlier resistivity and magnetization experiments on
Ga$_{84}^{4-}$ samples \cite{Hagel02}, indications were already
found for metallic behavior, and even superconductivity below a
transition temperature $T_{\text{c}}\approx 7.2$~K, much higher
than $T_{\text{c}}\approx 1.1$~K known for bulk $\alpha$-Ga metal.
The observed superconducting (SC) fraction was only 0.01\%
however, and, since it is known that Ga-metal in confined
geometries tends to have higher $T_{\text{c}}$ values, even as
high as 6~K \cite{Charnaya98}, these data were interpreted with
caution and an explanation in terms of bulk metal inclusions,
originating from possible deterioration and subsequent coalescence
of clusters, could not be ruled out \footnote{The Ga$_{84}$
compounds are highly air sensitive.}. Additional experiments were
needed therefore to provide unambiguous proof for the occurrence
of bulk superconductivity related to weak intermolecular charge
transfer, similar as in doped C$_{60}$ or other molecular
(super)conductors. Such proof is presented here on basis of
$^{69,71}$Ga-NMR experiments in both the metallic conducting and
the SC phases of these cluster solids.

\begin{figure}[tbp!] \center
\includegraphics[width=1\linewidth]{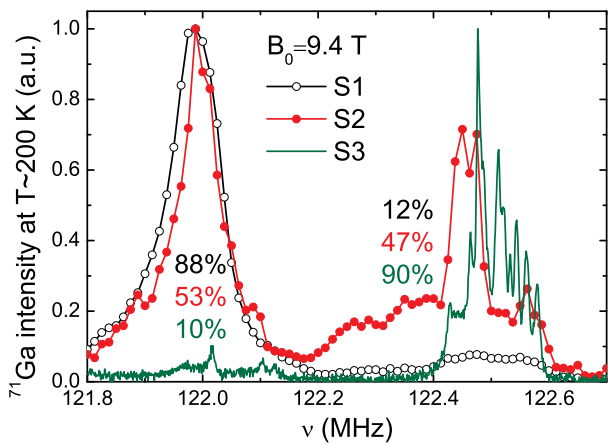}
      \caption{ \label{FigNC} High-$T$ spectra for
      S1, S2 and S3, corresponding to an unshifted reference frequency $\nu_{0}\approx 122~$MHz ($B_{0}\approx 9.4$~T).
      They show two main lines, with varying relative weights.
      }
      \end{figure}

NMR is an element-specific probe that can evidence two hallmarks
of metallic conductivity. The magnetic hyperfine interaction
between nuclear moments and conduction electron spins produces a
shift $\delta\nu$ of the NMR frequency $\nu_{0}$. This Knight
shift, $K_{S}=\delta\nu/\nu_{0}$, is proportional to the density
of states at the Fermi-energy, $D(E_{F})$. Secondly, this
interaction leads to a linear $T$-dependence (Korringa-law) of the
nuclear spin-lattice relaxation rate, $T_{1}^{-1}\propto aT$. The
proportionality constant is $a = K_{S}^{2}/S$, where $S$ is the
Korringa constant. For free $s$-electrons, $S$ is given
theoretically by $\mu_{B}^{2}/\pi\hbar k_{B} \gamma^{2}\approx
2.826\times 10^{-6}$~sK for $^{71}$Ga ($\gamma$ denotes the
nuclear gyromagnetic ratio). Moreover, NMR can also probe the
superconductivity: when below $T_{\text{c}}$ the conduction
electrons become spin-paired, both $T_{1}^{-1}$ and $K_{S}$ should
decrease. Experimental details about the $^{69,71}$Ga-NMR
technique can be found in a previous brief note \cite{Bakharev03}
on preliminary results for Ga$_{84}^{4-}$ samples in the metallic
conducting region ($T >10$~K). Meanwhile several samples of both
moieties have been studied and we present here representative
results on three samples, labelled S1 (Ga$_{84}^{4-}$), S2
(Ga$_{84}^{3-}$) and S3 (Ga$_{84}^{4-}$). High-$T$ ($\sim 200~$K)
$^{71}$Ga spectra are shown in Fig.~\ref{FigNC} and display two
main lines, one being unshifted and the other shifted by
$0.4-0.5$~MHz. The relative weight of these lines is sample
dependent. Calibration of the Ga-NMR signals with an Al reference
confirms, however, that the observed concentration of Ga nuclei in
the combined spectra is as expected from the chemical formulae of
the cluster compounds.

\begin{figure}[tbp!] \center
\includegraphics[width=1\linewidth]{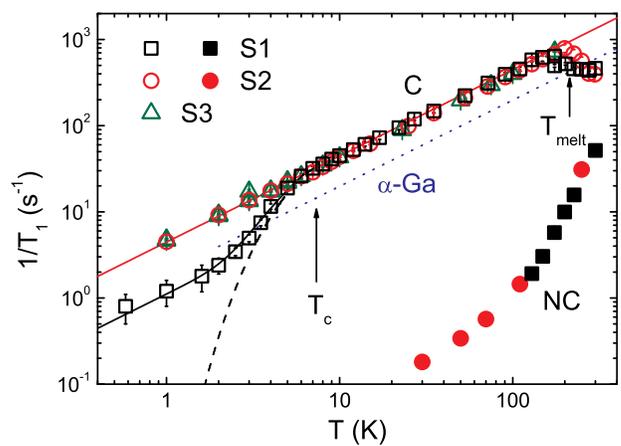}
      \caption{ \label{FigT1} $T$-dependence of the $^{71}$Ga-$T_{1}^{-1}$
      measured on the maxima of the two lines
      displayed in Fig.~\ref{FigNC}. The open symbols follow the
      Korringa law when $T\geq T_{\text{c}}$ (and $B_{0}\geq B_{\text{c2}}$),
      and refer to the C fraction
      (several fields $1.5~\text{T}\lesssim B_{0}\leq 9.4~$T
      yield the same Korringa constant). The SC transition at $T_{\text{c}}\approx7.5$~K is observed in
      S1 in $B_{0}\approx 2.39~$T.
      The dotted curve indicates the behavior of bulk $\alpha$-Ga.
      The other curves are fits (see text).
      Closed symbols refer to the unshifted line (NC
      fraction).
      }
      \end{figure}

Representative $T_{1}^{-1}$ data, measured on the maximum of the
lines, are given in Fig.~\ref{FigT1}. The closed symbols refer to
the unshifted line. The relaxation rate of these Ga-nuclei is one
to two orders of magnitude smaller than in $\alpha$-Ga metal
\cite{Hammond66} and does not follow the Korringa law. Hence, they
lack the relaxation channel provided by conduction electron spins
and only couple energetically to the phonons by quadrupolar
relaxation, as proved by the measured ratio $^{69}T_{1}/^{71}T_{1}
\approx (^{71}Q/^{69}Q)^{2} \approx 0.4$ (Q denotes the nuclear
quadrupolar moment). The unshifted line is therefore ascribed to a
``non-conducting'' (NC) fraction of the sample. By contrast, the
points measured on the shifted line (open symbols) obey very well
the Korringa law $T_{1}^{-1}\propto aT$ in a wide $T$-range (below
$T\sim 200~$K down to $T_{\text{c}}$), with $a =
4.45(20)$~s$^{-1}$K$^{-1}$ for all samples (i.e. independent of
the NC fraction). This line can thus be attributed to a conducting
(C) phase of the sample.

The origin of the NC fraction turns out to be a lack of toluene
molecules, normally present as a crystal solvent and apparently
essential to produce an \emph{ordered} 3D packing of the cluster
molecules. We found that when the crystals are taken out of the
toluene solution, toluene molecules rapidly diffuse out of the
sample which thereby looses its perfect crystalline order, with
accompanying loss of conducting properties. Indeed, as is well
known, in case of a small (intermolecular) charge transfer,
band-conductivity can be rapidly destroyed by even a small amount
of disorder (Anderson localization). In the investigated samples
the NC fraction varies from 88\% down to 10\% for the ``purest''
sample (S3, kept continuously in toluene liquid). Importantly,
independent of the C/NC ratio, we find always the same types of
behavior of $T_{1}$ for the NC fractions, as well as for the C
fractions in the metallic conducting region ($T > 8$~K). This is a
strong indication that the two phases exist separately, i.e. they
are not mixed on a molecular level in the material. Most likely
the NC phase is predominantly present at the surfaces of the
crystallites, where toluene as crystal solvent is most readily
removed. In the following, we only discuss the C fractions of the
samples.

\begin{figure}[tbp!] \center
\includegraphics[width=1\linewidth]{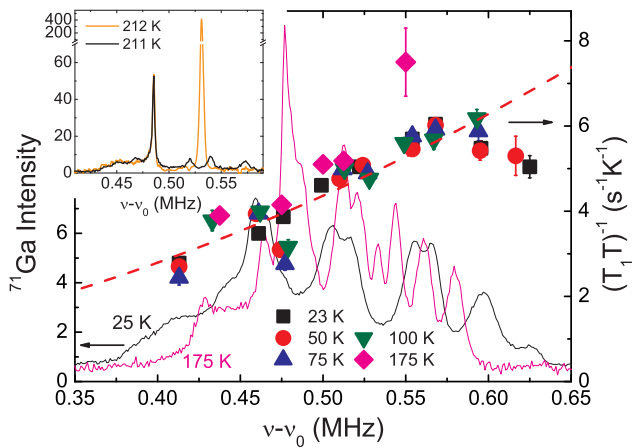}
      \caption{ \label{FigT1III} Left axis (lines) and inset: $T$-dependence of the $^{71}$Ga spectrum
      of the C fraction in S3 ($\nu_{0}\approx 122~$MHz).
      The motional narrowing of the lines is very pronounced
      above $T_{\text{melt}}\approx212$~K (inset) and is more gradual below.
      Right axis (closed symbols): frequency dependence of
      $(T_{1}T)^{-1}$ over the $^{71}$Ga spectrum in S3,
      for $T_{\text{c}}<T<T_{\text{melt}}$.
      The dashed line is a fit (see text).
      }
      \end{figure}

Interestingly, the value $a$ found for the C fraction is about 2.3
times that of bulk $\alpha$-Ga metal \cite{Hammond66}, indicating
a larger $D(E_{F})$ for the cluster compound. The value found for
the ratio of the $^{71}$Ga and $^{69}$Ga rates,
$^{69}T_{1}/^{71}T_{1} \approx (^{71}\gamma/^{69}\gamma)^{2}
\approx 1.6$ confirms the magnetic origin (the coupling to the
electron spin) of the nuclear relaxation. The $^{71}$Ga Knight
shift is \textbf{$^{71}K_{S}=0.39(3)$\%} for this line, using for
calibration the metallic $^{63}$Cu reference signal
($^{63}K_{S}=0.238$\%). We derive the Korringa constant,
$S=K_{S}^{2}/a= 3.4(6)\times10^{-6}$~sK. This is only $\sim$15\%
higher than the theoretical (free-electron) value, indicating a
predominantly $s$-like character of the itinerant electron
density, with weak correlations. For comparison, for bulk
$\alpha$-Ga the shift is $^{71}K_{S}\approx 0.16$\%, consistent
with a smaller $D(E_{F})$, and the resulting Korringa constant is
$\sim60$\% smaller than for the free electron case. A more
accurate method to evaluate $S$ is presented in
Fig.~\ref{FigT1III}, showing for several temperatures the
variation of $T_{1}T$ over the C-line in the $^{71}$Ga spectrum
for the (purest) sample S3. Despite the complexity of the spectrum
(see below), all data points are in excellent agreement with
metallic behavior with the same Korringa constant. A fit
$(T_{1}T)^{-1}=\frac{(\nu-\nu_{0})^{2}}{S\nu_{0}^{2}}$ is well
obeyed with $S=3.85(9)\times 10^{-6}$~sK. This is in good
agreement with the former result based on measurements of the
$T$-dependence of $T_{1}$ on the maximum of the line.

Another interesting feature is the strong effect of motional
narrowing observed in the $T$-dependence of the spectra, also
shown in Fig.~\ref{FigT1III}. The spectrum at low $T$ is $\sim
200$~kHz wide, with a fine structure that we attribute to the
different Ga-sites (with different interatomic distances) in the
cluster cores \cite{Schnepf01,Schnepf03}. For S3 (highest C
fraction), the spectrum can be decomposed into at least 10
different lines. Within a very narrow range of temperature ($\sim
1$~K), the whole spectrum collapses above $T_{\text{melt}}\approx
212~$K into two narrow lines, at 122.486 and 122.533~MHz
[Fig.~\ref{FigT1III}(inset)], the upper one having a linewidth at
room-$T$ of only 1~kHz in a field of 9.4~T. The motional rate
estimated from the linewidths increases from 0.4~MHz around
$T_{\text{melt}}$ to 10~MHz at room-$T$. Moreover, this line
contains $\sim70$\% of the Ga nuclei. Considering that $\sim 25$\%
of the Ga atoms form the nearest-neighbor shell around the central
Ga$_{2}$ dimer \cite{Schnepf01,Schnepf03}, and that X-ray studies
show the freezing of rotations of the Ga$_{2}$ dimer within this
cage in this $T$-range, we associate the lower frequency line to
the nearest-neighbor sites of the central dimer. These would be
most affected by the dimer motions, explaining the weak motional
narrowing observed even below $T_{\text{melt}}$ in particular for
this subspectrum (Fig.~\ref{FigT1III}). As seen in
Fig.~\ref{FigT1}, the transition at $T_{\text{melt}}$ is
accompanied by a drop (factor 3) in $T_{1}^{-1}$ of the C
fraction. Apparently, the onset of the internal motions affects in
some as yet unknown way the $D(E_{F})$ value.

The combination of all the above results leaves no doubt that the
observed spectra are due to Ga-nuclei belonging to the cluster
molecules and not to inclusions/particles of bulk Ga-metal. The
$T$-dependence of $T_{1}^{-1}$ is typical for nuclear relaxation
via conduction electrons. The fact that the conductivity is lost
when the ordered packing of the clusters is disturbed proves that
it arises from intercluster charge transfer.

      \begin{figure}[tbp!] \center
\includegraphics[width=1\linewidth]{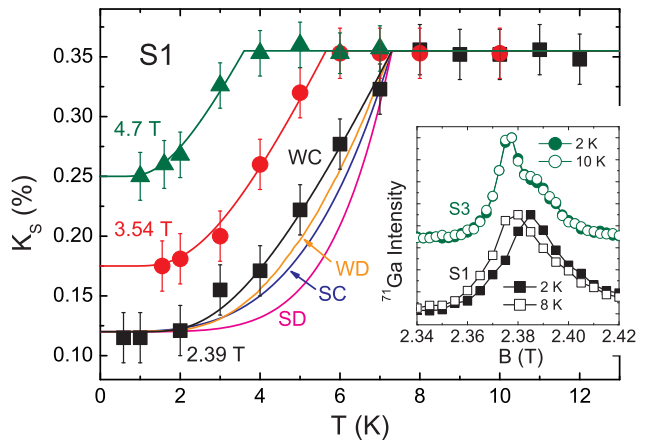}
      \caption{ \label{FigShift} $T$- and field-dependences of
      the Knight shift $K_{S}$ for S1.
      Data for lowest field are
      compared with BCS theory for weak and strong coupling
      (symbols ``W'' and ``S'') in the clean and dirty limits
      (symbols ``C'' and ``D''). The ``WC'' case is included for the other fields.
      Inset: $^{71}$Ga spectra at low-$T$.
      No shift in found for S3, having $B_{\text{c2}}\sim 0.25~\text{T}\ll B_{0}\approx 2.39~$T.
      }
      \end{figure}

Turning next to the SC properties, we remark that for all samples,
magnetization data show the occurrence of bulk superconductivity
of type II, with $T_{\text{c}}\approx 7.4~$K, 7.8~K and 7.95~K for
S1, S2 and S3, and a lower critical field $B_{\text{c1}}$ of the
order of a few 0.01~T. The upper critical field $B_{\text{c2}}$
appears to vary strongly, between $\sim $0.3~T (S2 and S3) up to a
few Tesla (S1), i.e. (very) much lower than the value of 13.8~T
reported in \cite{Hagel02}. In fact, these low values for
$B_{\text{c2}}$ are the reason why by NMR we could only detect
superconductivity in samples with sufficiently high
$B_{\text{c2}}$, since sensitivity requirements limited our NMR
experiments to $B_{0}\gtrsim 2~$T. As an example, the $T$- and
field-dependencies of $T_{1}^{-1}$ and $K_{S}$ are shown in
Figs.~\ref{FigT1} and~\ref{FigShift} for the C fraction of S1 for
$T\lesssim10~$K. They provide definite proof for the occurrence of
bulk superconductivity below $\approx 7.5~$K for this sample, up
to several Tesla. $K_{S}(T)$ can be well fitted with the
prediction from the BCS model with $s$-wave symmetry and weak
electron-phonon coupling, implying a SC gap
$\Delta/k_{B}=1.76\,T_{\text{c}}\sim13$~K. Several other samples
showed similar dependencies of the shift. The data in
Fig.~\ref{FigShift} do not allow to distinguish between the
predictions for the ``clean'' and ``dirty'' limits for the weak
coupling model, but the strong coupling case is clearly excluded.
We note that in Fig.~\ref{FigT1} we do not observe a coherence
peak in the $T$-dependence of $T_{1}^{-1}$ around $T_{\text{c}}$.
As recently discussed in \cite{Magishi05}, a strongly reduced
coherence peak in $T_{1}^{-1}$, in combination with a gapped
behavior (sharp drop) at a lower temperature than $T_{\text{c}}$,
is typical for an $s$-wave superconductor. As shown by the
continuous curve in Fig.~\ref{FigT1}, our $T_{1}$ data can indeed
be fitted, for $T\lesssim 0.85T_{\text{c}}$ by the sum of an
exponential ($\propto\exp(-\Delta/k_{B}T)$, dashed curve) with
$\Delta/k_{B}=13\pm2$~K and a linear term, accounting for the
relaxation of the normal electrons in the vortex cores
\cite{Magishi05}.

In view of the strong effect of the toluene molecules on the
conductivity, we attribute the strong sample dependence of
$B_{\text{c2}}$ to a varying degree of lattice defects, related to
missing toluene molecules in the otherwise ordered lattices of the
conducting Ga$_{84}$ phases. We recall that theoretical models for
``dirty'' superconductors predict a $B_{\text{c2}}$ inversely
proportional to the electronic mean free path \cite{Tinkham},
whereas the values of $T_{\text{c}}$ and the SC energy gap
$\Delta$ should not be affected much by even high concentrations
of defects (provided these are nonmagnetic), as observed. Though
we expect the NC fractions to be present as a separate (surface)
phase, a higher fraction of the (non-crystalline) NC phase may
well entail a higher amount of lattice defects in the
(crystalline) C phase of the sample. Indeed, S1 had the highest
amount ($\sim88$\%) of the NC phase \footnote{It was studied
before the problem of the missing toluene molecules was realized,
and had been taken out of the toluene solution.}. We expect the
sample studied in \cite{Hagel02}, with a reported
$B_{\text{c2}}\approx13.8$~T, to have contained an even larger
concentration of defects. As a check on this interpretation, Muon
Spin Relaxation experiments were performed at the PSI facility in
Switzerland on S3 in fields from zero to 0.3~T and $T>2~$K. The
results will be reported elsewhere \cite{Bono}, but we mention
here that also these data clearly prove the presence of bulk
superconductivity with $B_{\text{c2}}\sim 0.25$~T and
$B_{\text{c1}}\sim 50~$mT. Using theoretical expressions for the
critical fields of a type II superconductor
\cite{Tinkham,Brandt03}, one obtains $\lambda\approx 80$~nm and
$\xi\approx 40$~nm for the London penetration depth and the SC
coherence length. Both values are much larger than the cluster
core ($\varnothing\approx$1.4~nm) and the average distance between
the core centers ($\approx$2.3~nm), in agreement with bulk
superconductivity for the \emph{crystalline array} of clusters.
The quotient $\kappa=\lambda/\xi\approx2$ indicates
superconductivity of type II.

Summarizing, the above experiments provide compelling evidences
for a band-type conductivity by weak intercluster charge transfer.
How this charge transfer process occurs and via which
intermediates is at present unknown, but it is clearly very
sensitive to small local changes in the intercluster packing. This
is reminiscent of the orientational disorder effects in C$_{60}$.
Comparing the molecular structure with that of C$_{60}$, the
intercluster transfer integral (i.e. the bandwidth) is expected to
be much smaller, of order $t \sim 1-10~$meV, considering that the
Ga$_{84}$ cluster cores are separated by surrounding ligand
shells. On the other hand, the on-site Coulomb interactions will
also be smaller in view of the larger cluster size. This may
explain the -at first sight surprising- result that we find no
evidence for strong electron correlation effects, in spite of the
very narrow bandwidth.

Our findings present several important challenges to theory, such
as the occurrence of bulk superconductivity with relatively high
$T_{\text{c}}$ at such small $t$ values, and the nature of the SC
pairing mechanism. Assuming the latter to be phonon-mediated, the
Ga$_{84}$ compound appears an almost perfect first experimental
realization of the theoretical model advanced by Friedel
\cite{Friedel92} shortly after the discovery of superconductivity
in fullerenes. He showed that for a \emph{crystalline ordered
array of identical metal nanoclusters} even a weak intercluster
charge transfer can yield a large $T_{\text{c}}$, provided the
degeneracy of molecular levels near $E_{F}$ is sufficiently large.
In this respect we mention that a larger value for $D(E_{F})$ than
for $\alpha$-Ga was obtained for the Ga$_{84}$ compound from
Density Functional calculations \cite{Frenzel04}. Quite recently,
the same idea of obtaining a high $T_{\text{c}}$ by increasing the
$D(E_{F})$ of the molecular-level-derived band through reduction
of the transfer integral has been proposed \cite{Hamel05} for a
novel hybrid compound, that would consist of C$_{60}$ molecules
embedded in a metal-organic framework serving to keep them at a
larger (tunable) distance.

This work is part of the research program of the ``Stichting FOM''
and is partially funded by the EC-RTN ``QuEMolNa'' (No.
MRTN-CT-2003-504880), the EC-Network of Excellence ``MAGMANet''
(No. 515767-2), the DFG-Centre of Functional Nanostructures
(Karlsruhe).

\end{document}